\documentclass[11pt,a4paper]{article}
\usepackage{epsfig,graphics,color,psfrag}

 \normalsize
\newcommand{\be}{\begin{equation}}\newcommand{\ee}{\end{equation}}
\newcommand{\bea}{\begin{eqnarray}}\newcommand{\eea}{\end{eqnarray}}
\newcommand{\bc}{\begin{center}}\newcommand{\ec}{\end{center}}
\newcommand{\bmini}{\begin{minipage}}\newcommand{\emini}{\end{minipage}}
\def\ep{\eta^{\prime}}\def\jp{J/\psi}
\def\lsim{\raise0.3ex\hbox{$\;<$\kern-0.75em\raise-1.1ex\hbox{$\sim\;$}}}
\def\gsim{\raise0.3ex\hbox{$\;>$\kern-0.75em\raise-1.1ex\hbox{$\sim\;$}}}
\def\note#1{\textcolor{black}{ #1}}
\def\la{\langle}\def\ra{\rangle}
\def\matrix#1#2{\left(\begin{array}{@{\,}#1@{\,}} #2 \end{array}\right)}
\def\mpi{m_{\pi}}\def\mk{m_{K}}\def\Mp{M_{\pi}}\def\Mk{M_{K}}
\def\tR{\tilde{R}}\def\td{\tilde{\delta}}\def\mzt{\tilde{m}_0}
\def\oder#1{${\mathcal{O}( #1 )}$}
\begin{document}
\thispagestyle{empty}
\rightline{UCL-IPT-04-21}
\rightline{\today }

\vspace{2cm}

\centerline{\LARGE  $\eta - \ep$ masses and mixing: 
a large $N_c$ reappraisal}

\vspace{2cm}
\centerline{\large  J.-M. G\'{e}rard\footnote{ e-mail address: gerard@fyma.ucl.ac.be } 
and E. Kou\footnote{ e-mail address: ekou@fyma.ucl.ac.be}}
\bigskip
\vspace{.5cm}
\centerline{
Institut de Physique Th\'{e}orique, Universit\'{e} catholique de Louvain,}
\centerline{ Chemin du Cyclotron-2, B-1348, Louvain-la-Neuve, Belgium}
 \vskip0.5truecm
\vspace*{1truecm}

\centerline{\large Abstract }
We reconcile the $1/N_c$ expansion with the observed $\eta - \ep$ mass spectrum. 
The chiral corrections introduced for that purpose are natural and consistent with the octet-singlet 
mixing angle $\theta=-(22\pm 1)^{\circ}$ extracted from phenomenology in the large $N_c$ limit.  
\newpage
\pagenumbering{arabic}
\section{Introduction}
There is, nowadays, a considerable interest in weak decay processes involving 
$\eta$ and $\ep$ mesons as final or intermediate states. New physics beyond 
the Standard Model is indeed regularly advocated to explain, for example, 
the unexpectedly large $B\to K\ep$ branching ratio \cite{KK}
or the sizable direct CP violation in $K\to \pi\pi$ amplitudes \cite{Silvestrini}. 
This might be legitimate if non-perturbative hadronic effects such as a gluonium component in 
$\ep$ or a destructive $\pi^0 - \eta$ mixing contribution were fully under control. 
As a matter of fact, the $q\bar{q}$-gluonium mixing would vanish \cite{Witten} and all 
the $\Delta S=1$ hadronic matrix elements could be factorized \cite{Buras} 
if the number of colors \note{$N_c$} turned out to be infinite... 

Interestingly enough, the large $N_c$ approximation \cite{tH} has been proven to provide 
a simple and quite successful theoretical framework for elucidating various 
non-perturbative phenomena in strong dynamics. 
For illustration, 
the Okubo-Zweig-Iizuka (OZI) rule, which qualitatively explains the $\rho -\pi$ suppression in 
$\phi$ decay and the $\omega -\rho^0$ mass difference, can be 
coherently derived in this approximation. 
The chiral symmetry breaking pattern is also understood from the observed mass spectrum 
of the pseudoscalar mesons. However, more recently, this useful framework has been 
challenged at the quantitative level due to its apparent failure to reproduce the 
well-measured $\eta$ and $\ep$ masses \cite{GI}. 

In this letter, we argue that reasonable chiral corrections alone may reproduce the 
$\eta$  and $\ep$ masses  in the large $N_c$ limit. 
A direct extraction of the octet-singlet mixing angle in this limit confirms 
the natural size of these corrections. 
\section{Georgi's mass inequality revisited}
In the strict large $N_c$ limit, only the color-singlet channel of the 
quark-antiquark planar interaction is attractive and QCD with three massless flavors 
($u, d, s$) exhibits chiral symmetry breaking \cite{CW}. 
The $U(3)_L\times U(3)_R$ chiral symmetry of the fundamental QCD theory is indeed 
spontaneously broken down to $U(3)_{L+R}$ such that a full nonet of Goldstone bosons 
\be
\pi =\sum_{a=0}^{8}\lambda_a\pi^a 
=\sqrt{2}\matrix{ccc}{\frac{\pi^0}{\sqrt{2}}+\frac{\eta_8}{\sqrt{6}}+\frac{\eta_0}{\sqrt{3}}&\pi^+&K^+\\
\pi^-&-\frac{\pi^0}{\sqrt{2}}+\frac{\eta_8}{\sqrt{6}}+\frac{\eta_0}{\sqrt{3}}&K^0\\
K^-&\overline{K}^0&-\sqrt{\frac{2}{3}}\eta_8+\frac{\eta_0}{\sqrt{3}}} 
\label{eq:v5-1}
\ee
is naively expected. To lowest order in derivatives, the effective Lagrangian for the 
corresponding unitary field 
\be
U(x)=\exp (i\frac{\sqrt{2}\pi}{f}) 
\ee
reads then 
\be
{\mathcal{L}}_{\infty}^{(p^2)}=\frac{f^2}{8}\la \partial_{\mu}U\partial^{\mu}U^{\dagger}\ra
 \label{LLO}
\ee
where $f$ is the weak decay constant of the pseudoscalar nonet. 
The bracket $\la \cdots \ra$ stands for the trace over light flavors. 
The other possible kinetic term  
\be
{\mathcal{L}}_{1/N_c}^{(p^2)}= \epsilon_1\frac{f^2}{8}
\la \partial_{\mu} U U^{\dagger}\ra\la \partial^{\mu} U^{\dagger} U\ra
\label{eq:v5-8}
\ee
contains two traces. Such a flavor structure 
necessarily arises from QCD Feynman diagrams with two quark loops and 
is therefore suppressed by one power of $1/N_c$. 

Explicit symmetry breaking terms have to be introduced to 
reproduce the observed mass spectrum of the light pseudoscalar mesons. 
Let us classify these breaking terms according to the momentum 
expansion in the large $N_c$ approximation. 
In other words, at each order in $p^2$, let us only retain the dominant 
term in the $1/N_c$ expansion. 

At leading order ($p^0$), the first non-trivial term only arises at the $1/N_c$ level and   
breaks the flavor singlet axial $U(1)_A$ symmetry \cite{Ven}:
\bea
\Delta \mathcal{L}_{1/N_c}^{(p^0)}&=&\frac{m_0^2}{4N_c}\frac{f^2}{8}
\la \ln U-\ln U^{\dagger}\ra ^2 \nonumber \\
&=&-\frac{1}{2}m_0^2\eta_0^2. 
\label{eq:v4-1}
\eea
This octet-singlet mass-splitting is responsible for  the large $\ep$ mass \cite{epmass}. 

At next-to-leading order $(p^2)$, the  single trace term accounts for 
the $SU(3)_{L+R}$ symmetry breaking: 
\be
\Delta{\mathcal{L}}_{\infty}^{(p^2)}=\frac{f^2}{8}r\la mU^{\dagger}+Um\ra . \label{eq:v4-2}
\ee
Disregarding here the possibility of a tiny T violation, we identify $m$ with 
 the real diagonal quark mass matrix  
\be
m=\matrix{ccc}{m_u&0&0 \\0& m_d &0\\ 0&0&  m_s}.  
\ee
In the large $N_c$ limit, the observed pseudoscalar spectrum unambiguously 
determines the quark mass ratios  \cite{JM-Leu}  since the chiral transformation 
$m\to m+\alpha (\det m)m^{-1}$  \cite{Kaplan}
violates the OZI rule. Let us therefore consider the realistic isospin limit 
$(m_u=m_d\equiv \hat{m} \ll m_s) $ and work in the quark basis  
 \be
\pi=\sqrt{2} \matrix{ccc}{u\bar{u}&u\bar{d}&u\bar{s}\\ d\bar{u}&d\bar{d}&d\bar{s}\\
s\bar{u}&s\bar{d}&s\bar{s}} 
\ee
which is proving  to be more convenient than the octet-singlet one 
(privileged by Eq. (\ref{eq:v5-1})) to 
discuss the $\eta$ and $\ep$ masses. In this basis, we obtain 
$\mpi^2=r\hat{m}$ and $\mk^2=\frac{r}{2}(m_s+\hat{m})$ for the charged mesons and the 
following  mass matrix for the neutral $(u\bar{u}, d\bar{d}, s\bar{s})$ states: 
\be
M^2
=\frac{m_0^2}{N_c}\matrix {ccc}{1&1&1\\1&1&1\\1&1&1+R}
+ \mpi^2 \matrix{ccc}{1&0&0\\ 0& 1&0 \\ 0&0&1}  \label{eq:2-5}
\ee
with 
\be
R=\frac{2N_c}{m_0^2}(\mk^2-\mpi^2). 
\ee
This matrix includes both the $U(1)_A$ and the $SU(3)_{L+R}$ symmetry breaking terms 
in the large $N_c$ limit. Its straightforward diagonalization leads to 
\bea
m_{\eta}^2&=&\frac{m_0^2}{6}(3+R-\sqrt{9-2R+R^2})+\mpi^2 \nonumber \\
m_{\ep}^2&=&\frac{m_0^2}{6}(3+R+\sqrt{9-2R+R^2})+\mpi^2 
\label{eq:v5-6}
\eea
for $N_c=3$. 
With this assignment for the mass eigenstates, we easily obtain 
\be
0\leq \frac{m^2_{\eta}-\mpi^2}{m^2_{\ep}-\mpi^2}
\leq \frac{3-\sqrt{3}}{3+\sqrt{3}}\simeq 0.268. 
\label{eq:2-7}
\ee
 The lower bound, reached in the $m_0^2\to \infty$ limit, corresponds to 
 the octet approximation with the Gell-Mann-Okubo mass relation for $\eta =\eta_8$. 
The upper bound, saturated  for $R=3$, is a simple generalization of 
Georgi's inequality \cite{GI} for $\hat{m}\neq 0$.  The latter  requires {\it at least} 15 \% 
corrections from higher order terms in the effective Lagrangian to be compatible with 
the physical $\pi, \eta$ and $\ep$  masses: 
\be
\left(\frac{m_{\eta}^2-m_{\pi}^2}{m_{\ep}^2-m_{\pi}^2}\right)_{exp.}=0.313. 
\label{eq:v5-3}
\ee

The $1/N_c$ mass corrections at the  order ($p^2$) in the momentum 
expansion have already been advocated in \cite{GI} and \cite{PERIS}. 
Here we would like to emphasize that the large $N_c$ limit at the order ($p^4$) 
in the momentum expansion may be enough to  reproduce the $\eta -\ep$ mass spectrum. 
In fact, this second possibility seems to be favored by the extraction of the $\eta -\ep$ 
mixing from phenomenology as well as by the observed $SU(3)$ splitting among 
weak decay constants. 
\section{Mixing angle from phenomenology}
In the large $N_c$ limit, 
the physical $\eta$ and $\ep$ states decouple from gluonium states and 
are slightly off from $\eta_8$ and $\eta_0$, 
respectively, due to the \oder{p^2} $SU(3)$ breaking term in Eq. (\ref{eq:v4-2}). 
Consequently, they are parameterized in terms of a single and small mixing angle $\theta$ 
\note{associated with the diagonalization of the two-by-two mass matrix}:  
\bea
\eta& =& \eta_8 \cos\theta -  \eta_0\sin\theta \nonumber \\
\ep&=& \eta_8\sin\theta+\eta_0\cos\theta .\label{eq:2-2-3}
\eea

At this level, a first estimate of this well-defined mixing angle can be obtained from Eq. (\ref{eq:2-5}):
\be
\theta=-\frac{1}{2}\tan^{-1} \left[\frac{2\sqrt{2}R }{9-R}\right]. \label{eq:2-0}
\ee
For $R$ going to infinity, $\theta$ is shifted by $\pi/2$ at the singular point $R=9$ 
such that  the corresponding renaming $\eta \to \ep$ and $\ep \to -\eta$ required by 
Eq.  (\ref{eq:2-2-3}) is compatible with our assignment in Eq. (\ref{eq:v5-6}). 
Indeed, in this rather formal  limit, we revive  the so-called $U(1)$ problem \cite{Wein} 
with $\theta=\theta_{\mbox{\tiny ideal}}\simeq +35^{\circ}$. 
For $R=1$, we get $\theta\simeq -10^{\circ}$ with a totally unrealistic $\eta -\ep$ mass ratio. 
For the optimal value $R=3$ (see Eq. (\ref{eq:2-7})), we obtain $\theta\simeq -27^{\circ}$. 
So, a more precise determination of $\theta$ clearly requires 
a better fit of the $\eta -\ep$ mass spectrum, or vice versa. 
Here we choose a phenomenological extraction of this angle  
in order to get an upper bound on the \oder{p^4} chiral corrections 
needed to reproduce Eq. (\ref{eq:v5-3}). 

The explicit breaking of the flavor singlet axial $U(1)_A$ symmetry manifests 
itself as an anomaly in the divergence of the associated current: 
\be
\note{(\partial^{\mu}J_{5\mu}^0)_{\mbox{anomaly}}}=
\frac{3\alpha_s}{4\pi}G^a_{\mu\nu}\tilde{G}_{a}^{\mu\nu}. \label{eq:1} 
\ee
At the effective level, we obtain from Eqs. (\ref{LLO}) and (\ref{eq:v4-1})
\be
\note{(\partial^{\mu}J_{5\mu}^0)_{\mbox{anomaly}}}=
-\sqrt{3}fm_0^2\eta_0 \label{eq:3-2-3} 
\ee
such that $\alpha_sG_{\mu\nu}\tilde{G}^{\mu\nu}$ is a clean probe of the singlet component 
$\eta_0$ in $\eta$ and $\ep$ for OZI-suppressed processes \cite{SVZ}. 

Let us consider the well-measured OZI-suppressed processes, 
$\jp\to\eta\gamma$ and $\jp\to\ep\gamma$  where the initial $c\bar{c}$ annihilates into 
one photon by emitting two gluons. From the definition of the mixing angle in Eq.   (\ref{eq:2-2-3}), 
the amplitude ratio of these two processes can be written as
\be
R_{\jp}=\frac{A(\jp\to\eta\gamma)}{A (\jp\to\ep\gamma)}=
\frac{\la 0 | \alpha_sG^a_{\mu\nu}\tilde{G}_{a}^{\mu\nu}|\eta \ra}{\la 0 | \alpha_sG^a_{\mu\nu}\tilde{G}_{a}^{\mu\nu}|\ep \ra} = -\tan\theta
\label{eq:3-0}
\ee
due to the relations in Eqs. (\ref{eq:1}) and (\ref{eq:3-2-3}). 
Using the current  experimental value 
$\Gamma(\jp\to\eta\gamma)/\Gamma(\jp\to\ep\gamma)=0.200\pm 0.023$  
\cite{PDG}, we obtain: 
\be
\theta^{\mbox{\tiny exp.}}=-(22\pm 1)^{\circ}.  
\label{eq:3-2-5}
\ee
We can now estimate $R$ from Eqs. (\ref{eq:2-0}) and (\ref{eq:3-2-5}). 
The corresponding mass ratio obtained from Eq. (\ref{eq:v5-6}):  
\be
R=2.3\pm 0.1\ \ \  \Rightarrow\ \ \ 
\left(\frac{m_{\eta}^2-\mpi^2}{m_{\ep}^2-\mpi^2}\right)\simeq 0.26 
\ee
indicates that the required corrections for the $\eta -\ep$ masses are  in fact less than 20 \%. 

We would like to emphasize that 
the phenomenological extraction of $\theta^{\mbox{\tiny exp.}}$ presented in this section 
remains valid as long as no further $U(1)_A$ anomalous term arises in the effective Lagrangian.  
This  turns out to be the case for our momentum expansion in the large $N_c$ limit 
in which all the dominant breaking terms are single traces,  except for the 
\oder{p^0} effective Lagrangian! 
\note{Combining Eqs. (\ref{eq:v5-6}) and (\ref{eq:2-0}) to eliminate the SU(3)-breaking 
parameter $R$, we may equally express Eq. (\ref{eq:3-0}) in terms of the 
{\it theoretical} $\eta-\ep$ masses: 
\be
R_{J/\psi}=\cot [\theta +\tan^{-1}\sqrt{2} ]\left(\frac{m_{\eta}^2-m_{\pi}^2}{m_{\ep}^2-m_{\pi}^2}\right). 
\label{eq:vrev-1}
\ee
This relation resembles the standard PCAC one (see e.g.  \cite{BFT}). 
Notice however,  that a misuse of the 
{\it physical} $\eta-\ep$ masses (see Eq. (\ref{eq:v5-3})) at this level would imply $\theta\simeq -17^\circ$ instead of Eq. (\ref{eq:3-2-5}).}
\note{As we will see, higher order terms in $p^2$ do modify Eq. (\ref{eq:vrev-1}) but {\it not} 
Eq. (\ref{eq:3-0}), such that the phenomenological value of the mixing angle given in Eq. (\ref{eq:3-2-5}) 
is consistently obtained. }
 \section{Chiral corrections}
 The symmetry breaking Lagrangian at \oder{p^4} admits three terms with a single trace 
 over flavors: 
 \be
\Delta{\mathcal{L}}_{\infty}^{(p^4)}\hspace*{-0.1cm}= 
\frac{f^2}{8}\hspace*{-0.1cm}\left[-\frac{r}{\Lambda^2}\la m\partial^2 U^{\dagger}  \ra
+\frac{r^2}{2\Lambda_1^2}\la mU^{\dagger}mU^{\dagger}\ra
+\frac{r}{2 \Lambda_2^2}\la  mU^{\dagger}\partial_{\mu}U\partial^{\mu}U^{\dagger}\ra
\right] +h.c.
\label{eq:3-1}
\ee
The first and third terms  
modify  the weak currents  and induce the $SU(3)$ splitting between $\pi$ and $K$ decay constants 
\cite{PCP}:  
\be 
\frac{f_K}{f_{\pi}}-1=(\mk^2-\mpi^2)\left(\frac{1}{\Lambda^2}+\frac{1}{2\Lambda_2^2}\right). 
\label{eq:4-3-0}
\ee
From the observed value  \note{$f_K/f_{\pi}= 1.22\pm 0.01$}, we conclude \note{then} that 
$\Lambda$ and $\Lambda_2$ have to be   around 1 GeV, 
 the expected scale for any cut-off of the QCD effective theory. The 20 \%
corrections needed for the $\eta -\ep$ masses are therefore just at hand! 
From the second and third terms in Eq. (\ref{eq:3-1}), 
we now obtain the following mass matrix for the neutral $(u\bar{u}, d\bar{d}, s\bar{s})$ states:  
\be
\tilde{M}^2=\frac{\tilde{m}_0^2}{3}\matrix{ccc}{1&1&1-\tilde{\delta}\\1&1&1-\tilde{\delta}\\ 
1-\tilde{\delta}&1-\tilde{\delta}&1+\tilde{R}-2\td}+\Mp^2 \matrix{ccc}{1&0&0\\ 0& 1& 0\\ 0&0& 1}. 
\label{eq:4-4}
\ee
A quick glance from Eq. (\ref{eq:2-5}) to Eq. (\ref{eq:4-4}) displays the crucial appearance of 
a new parameter 
\be
\tilde{\delta}=\frac{\Mk^2-\Mp^2}{\Lambda_2^2} 
\label{eq:v6-1}
\ee
beyond the simple redefinition of $R$ and $m_0^2$:  
\bea
&\tilde{R}=\frac{6(\Mk^2-\Mp^2)}{\tilde{m}_0^2}
\left[1+(\Mk^2-\Mp^2)(\frac{2}{\Lambda_1^2}-\frac{1}{\Lambda_2^2})\right]&
\label{eq:4-5}\\
&\tilde{m}_0^2=m_0^2\left(1-\frac{\Mp^2}{\Lambda_2^2}\right).& \label{eq:4-3-5} 
\eea
Note that $\Mp$ and $\Mk$ stand now for the physical pion and kaon masses 
at ${\mathcal O}(p^4)$: 
\be
\Mp^2=\mpi^2\left[1+\mpi^2(\frac{2}{\Lambda_1^2}-\frac{1}{\Lambda_2^2})\right], \ \ \ 
\Mk^2=\mk^2\left[1+\mk^2(\frac{2}{\Lambda_1^2}-\frac{1}{\Lambda_2^2})\right]. 
\label{eq:4-6}
\ee
From the diagonalization of 
the mass matrix  in Eq. (\ref{eq:4-4}), we obtain  
\bea
\frac{M^2_{\eta}-\Mp^2}{M^2_{\ep}-\Mp^2}&=&\frac{3+\tR-2\td -
\sqrt{9-2\tR+\tR^2-4\tR\td-12\td +12\td^2}}
{3+\tR-2\td +\sqrt{9-2\tR+\tR^2-4\tR\td -12\td+12\td^2}} \nonumber \\
&& \label{eq:4-7}\\
M^2_{\eta}+M^2_{\ep}-2\Mp^2&=& \frac{\mzt^2}{3}(3+\tR-2\td)  \label{eq:4-8}\\
\theta&=& -\frac{1}{2}\tan^{-1} \left[\frac{2\sqrt{2}(\tR -3\td)}{9-\tR-6\td}\right]. 
\label{eq:4-9}
\eea

\begin{figure}[t]\bc
\psfrag{x}[c][c][1.2]{$\tR$}\psfrag{y}[c][c][1.2]{$\td$}
\psfrag{0.05}[c][c][1]{0.05}\psfrag{0.1}[c][c][1]{0.10}\psfrag{0.15}[c][c][1]{0.15}
\psfrag{0.2}[c][c][1]{0.20}\psfrag{0.25}[c][c][1]{0.25}
\psfrag{1.6}[c][c][1]{1.6}\psfrag{1.8}[c][c][1]{\ }\psfrag{2}[c][c][1]{2.0}
\psfrag{2.2}[c][c][1]{\ }\psfrag{2.4}[c][c][1]{2.4}\psfrag{2.6}[c][c][1]{\ }  \psfrag{2.8}[c][c][1]{2.8}\psfrag{3}[c][c][1]{\ }
\psfrag{3.2}[c][c][1]{3.2} \psfrag{3.6}[c][c][1]{3.6}
\psfrag{mass}[c][c][1]{masses}
\psfrag{mixing}[c][c][1]{\rotatebox{80}{mixing}}
\psfrag{c}[c][c][1.1]{$\bullet$}\psfrag{s}[c][c][1.25]{$\star$}
\includegraphics[width=10cm]{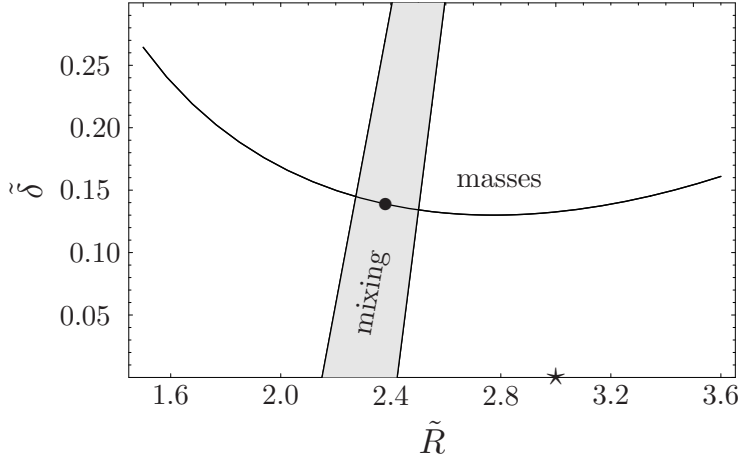}
\caption{Fit of the  \oder{p^4} chiral corrections using 
the observed masses and mixing angle for $\eta$ and $\ep$. 
The line slowly curving around $\tR=3$  is obtained from  Eqs. 
(\ref{eq:v5-3}) and (\ref{eq:4-7})  while 
the allowed shaded area is obtained from Eqs. (\ref{eq:3-2-5}) and (\ref{eq:4-9}) 
(left and right bounds correspond to $\theta=-21^{\circ}$ and $\theta=-23^{\circ}$, respectively). 
The black dot leads to Eq. (\ref{eq:v6-10}). The star indicates the \oder{p^2} 
optimal fit \cite{GI}  with $\theta \simeq -27^{\circ}$. 
} 
\label{fig1}
\ec\end{figure}

Using the observed values for the pseudoscalar masses $M_{\eta, \ep}^2$ and  the mixing angle 
$\theta^{\tiny exp.}$ (see Eq. (\ref{eq:3-2-5})),  
we can fix  the parameters $\tR, \mzt$ and $\td$.  Our main result is shown in Fig. \ref{fig1}. 
The horizontal line slowly curving around $\tR=3$ and 
the  quasi vertical lines are obtained from Eq. (\ref{eq:4-7}) and  Eq. (\ref{eq:4-9}), respectively. 
A remarkable property is that the constraints from  the mass ratio 
and the mixing angle are quite independent of 
$\td$ and $\tR$, respectively, at their intersection. 
Therefore, as the experimental errors for the 
radiative $\jp$ decays are small,  $\td$ and $\tR$ are rather  
precisely determined from this complementary analysis. Using the central value for the mixing as well as 
Eq. (\ref{eq:4-8}), we obtain the reasonable values for the three parameters in $\tilde{M}^2$:
\be
\td \simeq 0.14, \ \ \ \tR \simeq 2.4, \ \ \ \mzt\simeq 0.83\   \mbox{GeV}. \label{eq:v6-10}
\ee
As a result, all the cut-off parameters in the \oder{p^4} Lagrangian  are fixed 
around 1 GeV; 
\be
\Lambda \simeq 1.2 \ \mbox{GeV},  \ \ \ \Lambda_1\simeq 1.2\ \mbox{GeV}, \ \ \ \Lambda_2\simeq 1.3\ \mbox{GeV} \label{eq:v6-13}
\ee
as it should be. Therefore, 
the masses and mixing can be  quite naturally reproduced in the large $N_c$ limit. 

\note{Note that the new, $\tR$-independent, identity derived now from Eqs. (\ref{eq:4-7}) and (\ref{eq:4-9}) reads 
\be
-\tan\theta = \cot [\theta +\tan^{-1} \sqrt{2}] \left(\frac{M_{\eta}^2-M_{\pi}^2-\frac{2}{3}\mzt^2\td}{M_{\ep}^2-M_{\pi}^2-\frac{2}{3}\mzt^2\td}\right). 
\ee
By analogy with Eq. (\ref{eq:vrev-1}), we may rewrite 
\be
R_{J/\psi}=-\tan\theta =\cot [\tilde{\theta}+\tan^{-1}\sqrt{2}] 
\left(\frac{M_{\eta}^2-M_{\pi}^2}{M_{\ep}^2-M_{\pi}^2}\right)
\ee
with $\theta \simeq -22^{\circ}$ but $\tilde{\theta}=-17^{\circ}$. This result 
confirms the need for a two-angle formalism \cite{KL,FKS}, once one goes beyond 
PCAC to derive electroweak decay amplitudes. }

\section{$1/N_c$ corrections}
\note{Loops as well as tree-level multi-traces over flavors provide the $1/N_c$ corrections. 
One-loop corrections to the ratio $f_K/f_\pi$ turn out to be numerically small if the renormalization scale associated with the chiral logarithms is chosen in the vicinity of the $\eta$ mass. The large $N_c$ limit adopted here legitimates this rule-of-thumb such that our }
successful understanding of the $\eta -\ep$ masses and mixing in the 
large $N_c$ limit is basically due to the $\Lambda_2$ term in $\Delta{\mathcal{L}}^{(p^4)}_{\infty}$. 
This term could in principle be rotated away via a specific  \oder{p^2} transformation 
\be
U\to U^{\prime}=U+\frac{r}{4\Lambda_2^2}[m-Um^{\dagger}U] \label{eq:v6-11}
\ee
on ${\mathcal{L}}^{(p^2)}_{\infty}$. Such a field redefinition preserves the unitarity of $U$ 
up to \oder{p^4}. It eliminates the $\Lambda_2$ term which causes 
tedious (and sometimes overlooked \cite{PERIS}) 
wave-function renormalizations and  simply amounts to the substitutions
\bea
\frac{1}{\Lambda^2}&\to&\frac{1}{\Lambda^2}+\frac{1}{2\Lambda_2^2} \\
\frac{1}{\Lambda_1^2}&\to&\frac{1}{\Lambda_1^2}-\frac{1}{2\Lambda_2^2}
\eea
in $\Delta{\mathcal{L}}^{(p^4)}_{\infty}$, as seen from Eqs.  (\ref{eq:4-3-0}) and (\ref{eq:4-5}), 
respectively.  However, acting simultaneously on 
$\Delta{\mathcal{L}}^{(p^0)}_{1/N_c}$, this chiral transformation 
would then require a $1/N_c$-suppressed double trace term 
in  the \oder{p^2} effective Lagrangian:  
\be
\Delta{\mathcal{L}}^{(p^2)}_{1/N_c}=\epsilon_2\frac{f^2 }{8}r
\la mU^{\dagger}-Um\ra\la \ln U-\ln U^{\dagger}\ra \label{eq:4-2-12}
\ee
to absorb its effect via another harmless substitution: 
\be
\epsilon_2 \to \epsilon_2+\frac{m_0^2}{12\Lambda_2^2}. \label{eq:v6-12}
\ee
The $\epsilon_2$ term being forbidden in our large $N_c$ limit, the field redefinition given in 
Eq. (\ref{eq:v6-11}) is not allowed. 
The physical effect of the  $\Lambda_2$ cut-off on the pseudoscalar mass matrix 
(Eq. (\ref{eq:4-4})) is therefore a direct consequence of the expansion adopted here. 

To summarize, 
the $1/N_c$ corrections to \oder{p^2} terms in Eqs. (\ref{eq:v5-8}) and (\ref{eq:4-2-12}) are 
assumed to be negligible in our approach based on the hierarchy 
\be
{\mathcal{O}}(p^2, 1/N_c) \ll {\mathcal{O}}(p^4, \infty). 
\ee
In an alternative combined expansion \cite{ncexp} in $p^2$=\oder{\delta} 
and $1/N_c$=\oder{\delta}, 
the $\epsilon_1$ term would imply a wave-function renormalization of the $\eta_0$ field 
and, consequently, 
a global rescaling of the $q\bar{q}$ mass matrix.  
The $\epsilon_2$ term considered in \cite{KL} is more problematic. Its contribution to the mass matrix could of course 
be absorbed into $\td$, $\tR$ and $\mzt$ (see Eq. (\ref{eq:v6-12})). 
But being not invariant under $U(1)_A$, 
it would definitely invalidate our phenomenological extraction of the mixing angle from OZI-suppressed 
processes. 
Consequently, while the constraint from masses displayed 
in Fig. \ref{fig1} would remain the same, 
the one from mixing would simply disappear. 
So, from this point of view, $1/N_c$ corrections are not only unnecessary in reproducing  
the $\eta$ and $\ep$ masses but they also generate an ambiguity 
for the size of the chiral corrections estimated in Eq. (\ref{eq:v6-13}). 
 \section{Conclusions}
We have shown that the \oder{p^2} prediction for the  $\eta$ and $\ep$ masses 
indeed requires $15 \sim 20$ \% of higher order corrections. 
In the large $N_c$ limit, 
the \oder{p^4} corrections, which are welcome to explain the $SU(3)$ splitting of 
the $f_K$ and $f_{\pi}$ weak decay constants, naturally fill up  this deficit if 
the  octet-singlet mixing angle $\theta=-(22\pm1)^{\circ}$ consistently extracted from 
$\jp$ OZI-suppressed decays is used.  
The large $N_c$ approximation at each order in 
the momentum expansion provides therefore a simple and coherent description of 
the $\eta -\ep$ mass spectrum and mixing. 
The $1/N_c$ expansion being trustworthy, we are now in a favorable position 
to constrain new physics from (electro-)weak processes involving $\eta$ and $\ep$ mesons. 
\vspace{0.8cm}

\noindent
{\bf \large Acknowledgments}\\
This work was supported by the Belgian
Federal Office for Scientific, Technical and Cultural Affairs through the
Interuniversity Attraction Pole P5/27. 

 \end{document}